\documentclass[aapm,mph,reprint,graphicx, onecolumn]{revtex4-1}
\usepackage{graphicx}
\usepackage{color}
\usepackage{amsmath}
\usepackage{amsfonts}
\usepackage{bm}
\usepackage{ulem}
\usepackage{wrapfig}

\newcommand{\beq}{\begin{eqnarray}}
\newcommand{\eeq}{\end{eqnarray}}

\usepackage{comment}
\usepackage{epstopdf}

\usepackage{setspace}

\newcommand{\argmax}{\mathop{\rm arg~max}\limits}

\begin{document}

\setstretch{2.0}

\title{Technical Note: Fast Statistical Iterative Reconstruction for MVCT in TomoTherapy}


\author{
Sho Ozaki$^1$, Akihiro Haga$^{2{\rm a}}$, Edward Chao$^3$, Calvin Maurer$^3$, Kanabu Nawa$^1$, Takeshi Ohta$^1$, Takahiro Nakamoto$^1$, Yuki Nozawa$^1$, Taiki Magome$^4$, Masahiro Nakano$^5$, Keiichi Nakagawa$^1$
\\
{\it $^1$Department of Radiology, University of Tokyo Hospital, Tokyo, Japan,}\\
{\it $^2$Graduate School of Biomedical Sciences, Tokushima University, Tokushima, Japan} \\
{\it $^3$Accuray Incorporated. Sunnyvale, CA, USA}\\
{\it $^4$Radiological Science, Komazawa University, Tokyo, Japan}\\
{\it $^5$Radiation Oncology Department, Cancer Institute Hospital, Japanese Foundation for Cancer Research, Tokyo, Japan} \\
\vspace{1cm}
{\bf Numbers of Text pages, Figures, and Tables} \\
Total pages: 11 (including title page), Tables: 1, Figures: 4 \\
\vspace{1cm}
{\bf Running head} \\
Fast Iterative Reconstruction in TomoTherapy \\
\vspace{1cm}
{\bf Contact information} \\
a) Electronic mail: haga@tokushima-u.ac.jp \\
Telephone number: +81 88 633 9024 \\
Fax number: +81 88 633 9024 \\
Mailing address: 3-18-15, Kuramoto, Tokushima, 770-8503, Japan
}

\maketitle

\clearpage


\noindent
{\bf Abstract} \\
{\bf Purpose:} 
Statistical iterative reconstruction is expected to improve the image quality of megavoltage computed tomography (MVCT).
However, one of the challenges of iterative reconstruction is its large computational cost. 
The purpose of this work is to develop a fast iterative reconstruction algorithm by combining several iterative techniques and by optimizing reconstruction parameters.
\\
{\bf Methods:} 
Megavolt projection data was acquired from a TomoTherapy system and reconstructed using our statistical iterative reconstruction.
Total variation was used as the regularization term and the weight of the regularization term was determined by evaluating signal-to-noise ratio (SNR), contrast-to-noise ratio (CNR), and visual assessment of spatial resolution using Gammex and Cheese phantoms.
Gradient decent with an adaptive convergence parameter, ordered subset expectation maximization (OSEM), and CPU/GPU parallelization were applied in order to accelerate the present reconstruction algorithm.
\\
{\bf Results:} 
The SNR and CNR of the iterative reconstruction were several times better than that of filtered back projection (FBP).
The GPU parallelization code combined with the OSEM algorithm reconstructed an image several hundred times faster than a CPU calculation.
With 500 iterations, which provided good convergence, our method produced a 512$\times$512 pixel image within a few seconds.
The image quality of the present algorithm was much better than that of FBP for patient data.
\\
{\bf Conclusion:} 
An image from the iterative reconstruction in TomoTherapy can be obtained within few seconds by fine-tuning the parameters.
The iterative reconstruction with GPU was fast enough for clinical use, and largely improve the MVCT images.


\section{Introduction}

Helical TomoTherapy (HT) (Accuray, Sunnyvale, CA) is an innovative machine that delivers intensity modulated radiation therapy (IMRT)~\cite{Mackie:1993, Mackie:2006}, and includes megavoltage computed tomography (MVCT) for image guided radiation therapy (IGRT)~\cite{Ruchala:1999, Meeks:2005}.
In order to ensure precise image guidance, the image quality of the MVCT is crucial.
In particular, detection of soft tissue contrast is very important for soft tissue based registration accuracy and the accuracy of adaptive radiotherapy depends on the accuracy of the soft tissue deformation during the daily treatment~\cite{Langen:2005}.
However, in general, MVCT images are noisier than that of kilovoltage CT (kVCT)~\cite{Ruchala:1999}.
Furthermore, the relative difference in attenuation coefficients of various soft tissues at MV energies is smaller than at kV energies.
The smaller soft tissue attenuation differences lead to smaller contrast differences in MVCT and make it more difficult to 
register soft tissues accurately.

Statistical iterative reconstruction can improve the image quality of MVCT without increasing the imaging dose.
Unlike conventional reconstruction methods such as the filtered back projection (FBP), the iterative reconstruction methods can take into account {\it{a priori}} information including the CT geometry, photon statistics at the detector, and known properties of the image, in addition to the observed projection data. With such knowledge-based information, the iterative reconstruction method can reduce noise and improve soft tissue contrast.
Accuray has introduced an iterative reconstruction algorithm, "CTrue IR", on their Radixact platform that appears to improve the MVCT image quality~\cite{AccurayURL}. However, there aren't any publications describing the Accuray's algorithm as well as any source codes available. 
This is one of the motivations of this paper despite the availability of the Accuray CTrue IR.

A disadvantage of the iterative reconstruction is the large computational cost.
Since the method iteratively calculates a reprojection to compare with measured projection data, the total calculation time can be large. 
In this note, we describe an acceleration of the statistical iterative reconstruction for TomoTherapy MVCT by utilizing a parallelized CPU/GPU calculation. 
There are several reports on GPU based iterative reconstruction for cone-beam CT (CBCT) ~\cite{Scherl: 2007, Jia:2010, Xu:2012}.
The reconstruction time for a set of CBCT images with the iterative reconstruction algorithm could be shortened to $77$ to $130$s 
(about $100$ times faster than with conventional approaches). 
In addition to the GPU parallelization, we applied several acceleration techniques including an adaptive convergence parameter for the gradient decent method and the ordered subset expectation maximization (OSEM) algorithm. The performance and the quality of the reconstructed images of our method were evaluated using phantom and patient data.

\section{Iterative reconstruction scheme}
In this section, we briefly explain the basics of the statistical iterative CT reconstruction. 
In this study, the maximum {\it a posteriori} (MAP) approach was employed as our image reconstruction framework\cite{Tang2009a, Hanson1983a, Green1990, Lange1995}.
The concept of the MAP approach is to maximize the posteriori probability of reconstruction images with observed projection (or sinogram) data and a prior information about the image, encoded in the regularization term.
Reconstructed images can be obtained from an iterative process to maximize the log {\it a posteriori} probability function $\ln P(\bm{\mu}^* | \bm{y} )$ as,
\begin{align}
	\bm{\mu}^{*} 
	& =  \argmax_{\bm{\mu}^{*}}
	\left[  
	\ln{P(\bm{\mu}^*|\bm{y})} 
	\right] \hspace{2mm}{\rm subject\hspace{1mm}to} \hspace{2mm} \bm{\mu}^{*} > \bm{0},
   \label{eq:expected_mu}
\end{align}
where
\begin{align}
	\ln P(\bm{\mu}^* | \bm{y} ) \sim  \ln P(\bm{y} | \bm{\mu}^* ) + \ln P( \bm{\mu}^*) 
	= L ({\bm{\mu}}^{*} ) + \alpha R ({\bm{\mu}}^{*}) .
   \label{eq:map}
\end{align}
Here $L ({\bm{\mu}}^{*} ) = \ln P(\bm{y} | \bm{\mu}^*)$ expresses the log-likelihood probability function of observing the projection data set $\bm{y}$ at the given expectation of the image $\bm{\mu}^*$, whereas $\alpha R ({\bm{\mu}}^{*} ) = \ln P(\bm{\mu}^*)$ stands for the regularization term in the optimization process with a hyper parameter $\alpha$ controlling the weight of the regularization. 
We assumed that the photon statistics at the detector obeys the Poisson distribution and employed the total variation (TV) as the regularization term. The TV term penalizes a large difference between the image values of a certain position and its neighbors, and thus reduces the noise of the image. At the same time, however, this term also reduces the 
sharpness of object edges in the image, which leads to edge blurring.
Therefore, in general, the benefit of noise reduction must be balanced against the negative impact of edge blurring.
We seek the appropriate value of the weight $\alpha$ by examining the trade-off.

 In the optimization of Eq.~(\ref{eq:expected_mu}), the gradient decent method including the parameter adaptation was employed,
\begin{align}
\mu_{r}^{* (n+1) } 
= \mu_{r}^{* (n)} + \lambda^{(n)} \frac{ \partial }{ \partial \mu_{r}^{* (n)} }  
\left\{ L ( \bm{\mu}^{*(n) } ) + \alpha R(\bm{\mu}^{* (n) }) \right\} ,
\label{gdm}
\end{align}
where $n$ is the number of the iteration step, and $r$ means the index of the pixel to be reconstructed. $\lambda^{(n)}$ is the parameter which is adjusted to accelerate the calculation iteration step by iteration step as described in the next section. It is noted that, as well as the value of $\lambda^{(n)}$, the initial image setting $\mu_{r}^{* (0)}$ can affect the convergence speed.

\section{Setups and rapid convergence methods}

\subsection{Data acquisition and reconstruction}

Gammex phantom (Gammex, Middleton, WI) and Cheese phantom (Accuray, Sunnyvale, CA) were used in the analysis as well as the two patients for visual demonstration. All the data were acquired at the University of Tokyo Hospital during 1st - 31st May, 2016.
The reconstruction size of the image is $512 \times 512$ with an equal pixel size of 1mm for the one slice.

\subsection{Computer used in reconstruction}

The CPU/GPU specs in the computer used in this study (HP customized workstation: Z840/Z800) are shown as follows:

\begin{itemize}
 \item CPU: Xeon(R) X5690 ($\times2$: $12$ threads) / Xeon(R) E5-2687Wv ($\times2$: $20$ threads)
 \item GPU: GeForce GTX 1080 Ti
\end{itemize}

\subsection{Parameter optimization}

\subsubsection{Total variation parameter}
The weight of the TV regularization term $\alpha$ was set as $0$, $0.0001$, $0.0002$, and $0.0003$. The optimal values of the weight parameter were determined by estimating the signal-to-noise ratio (SNR), the contrast-to-noise ratio (CNR), the visual resolution, and the edge blurring effect. 
For SNR/CNR and edge blurring analyses, the Gammex phantom was used, whereas for visual resolution, the Cheese phantom was used.
The region-of-interests (ROIs) used in the SNR/CNR analysis as well as their definitions and the edge blurring analysis method
are mentioned in the supplemental material. 

\subsubsection{Convergence parameter and initial image setting}
Although a relatively large value of $\lambda^{(n)}$ in Eq.(\ref{gdm}) gives a rapid convergence, a too large value leads to non-convergence. 
Also the best value could depend on the iteration step. Therefore, we employed the following adaptive convergence parameter which depends on the iteration step: when the objective function being the square of difference between a reprojection and a projection decreases, the convergence parameter is increased as
\begin{align}
\lambda^{(n)}
= \lambda^{(n-1)} + 0.0001.
\end{align}
When the objective function increases, the parameter is decreased as
\begin{align}
\lambda^{(n)}
= \lambda^{(n-1)} - 0.0002.
\end{align}
The initial value of this parameter was set as 0.01.

As mentioned, the initial image setting $\mu_{r}^{* (0)}$ can also affect the convergence speed. 
In this study, homogeneous images with $\mu_{r}^{* (0)} = 0$, $0.05$, and $0.1$ [cm$^{-1}$] were tested.

\subsection{Ordered subset expectation maximization}

The ordered subset expectation maximization (OSEM) is a well-known method to accelerate the convergence of the iterative reconstruction algorithm~\cite{Hudson:1994}.
A basic idea of OSEM is to divide the original projection data into a subset of the data and to use the different subset in each iteration step. The easiest way to divide the projection data into subset is to thin the sinogram height out.
In the HT system, the data sampling time is $80$ Hz, and it takes $10$ sec for one rotation.
Hence the sinogram height is $800$. We thinned the sinogram height out by $1/2$, $1/5$, $1/10$, $1/20$, and $1/40$ (defined as the thinning parameter) with an equal interval (the sinogram height used in each iteration step is $400$, $160$, $80$, $40$, and $20$, respectively). For instance, in the $1/40$ case, $40$ iterations are needed to cover the whole sinogram data, so that the convergence could be relatively slow. Therefore, the convergence behavior as well as the convergence efficiency has to be verified carefully.
Here, it should be noted that the regularization weight $\alpha$ in the OSEM method is multiplied by the thinning parameter in order to keep the ratio of log-likelihood function with regularization term.

\subsection{Parallel computing}

For the parallel computing, we applied openMP for CPU parallelization, and cuda for GPU parallelization.
We have made our source code  available at https:$\cdots$.
In the next section, the reconstruction speeds with the several cases of the number of thinning out for OSEM described above will be compared.

\begin{figure}
\begin{center}
\includegraphics[width=12cm]{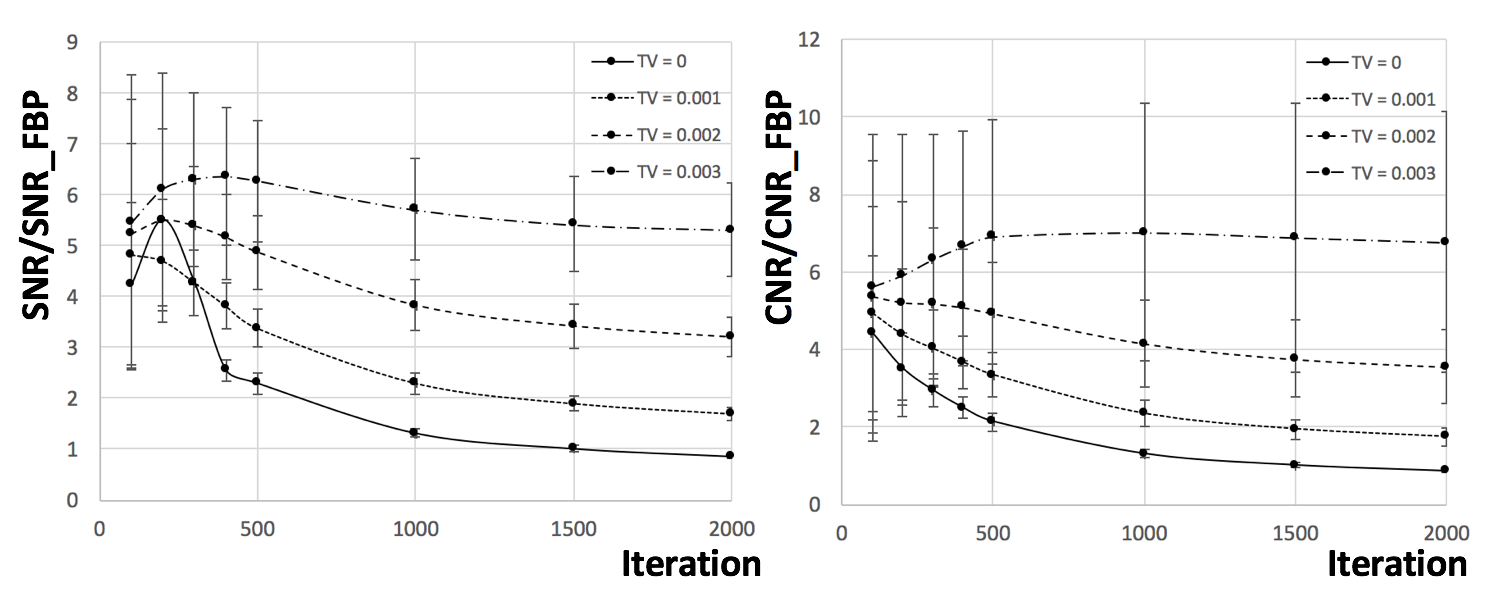}
\end{center}
\caption{SNR and CNR.}
\label{fig1}
\end{figure}

\section{Results and discussion}

\subsection{Total variation parameter}

\begin{figure*}[t]
\begin{center}
\includegraphics[width=11cm]{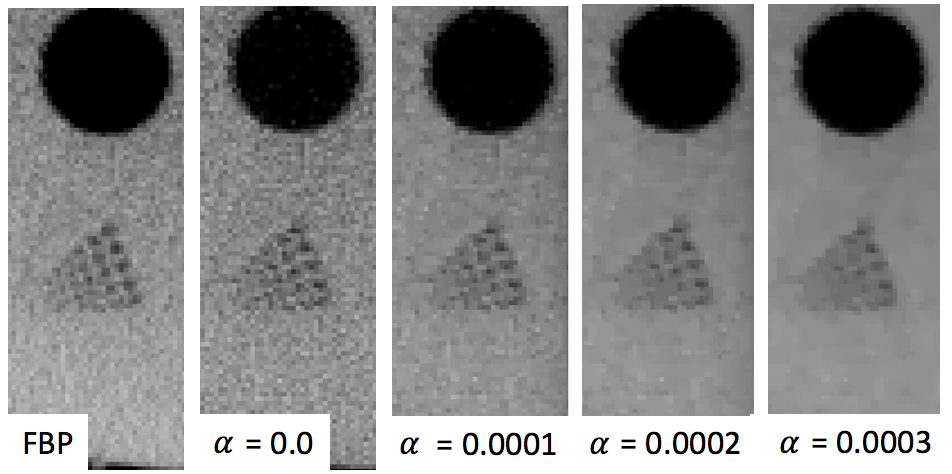}
\end{center}
\caption{Visual assessment for spatial resolutions (Cheese phantom).}
\label{fig2}
\end{figure*}

In order to determine the total variation parameter, the SNR and the CNR in the eleven insertions of the Gammex phantom were analyzed (see Fig.~S-2 in the supplemental material for the reconstructed images).
Figure 1 shows the results of the SNR and the CNR. In these results, the SNR and the CNR of the iterative reconstruction are normalized by the values of the FBP. 
In Fig.~1, both of the SNR and the CNR converge as the iteration number increases, and their values with the finite $\alpha$ are always larger than that of the FBP image.
On the other hand, as mentioned above, noise suppression is generally accompanied by a trade-off of reduced image sharpness.
This is seen in Fig.~2, where we can see that as the total variation parameter increases, the noise of the image decreases but the visibility of the small holes in the resolution plug of the cheese phantom also decreases.
From these results, we concluded that $0.0001 \le \alpha \le 0.0002$ maintains acceptable image resolution while achieving significant noise suppression.

\begin{figure}
\begin{center}
\includegraphics[width=9.2cm]{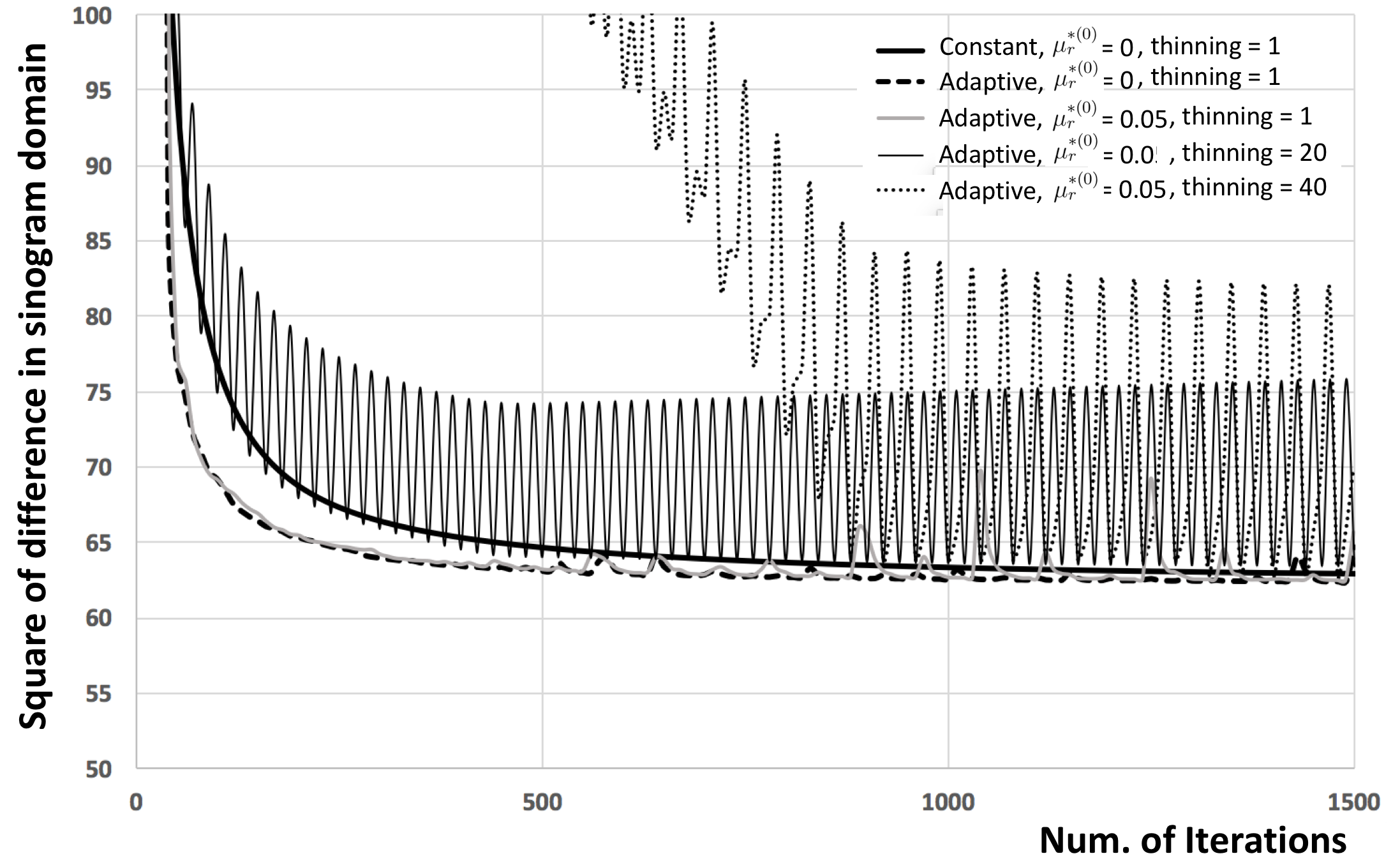}
\end{center}
\caption{Convergence behavior as a function of the number of iterations.}
\label{fig4}
\end{figure}

\subsection{Convergence parameter and initial image setting}

Figure 3 shows the convergence behavior for a variety of parameters, where convergence is characterized by
the square of the difference between a projection and a reprojection in the sinogram domain.
As expected, better convergence is obtained with the adaptive convergence parameter method.
With the adaptive convergence method, a similar degree of convergence is achieved with 500 iterations compared to 1500 iterations using a constant convergence parameter.
It should be noted that
the oscillatory behavior of the square difference is also caused by this adaptation; 
As the convergence parameter becomes gradually large according to the Eq.~(4), the images grow away from the optimal values. Eventually, the square difference turns to increase and numerically diverge.
This increasing behavior is however suppressed by decreasing the parameter as Eq.~(5) when the square difference is increased.
Due to this oscillation, the reconstructed image at the minimum of the oscillatory behavior of the square difference was employed.
The initial homogeneous images were set to values of $0$, $0.05$ and $0.1$ [cm$^{-1}$], and the best convergence was obtained for $0.05$ [cm$^{-1}$] in the Gammex phantom, though the difference was so small.
This was an expected result because the reconstructed value of $\mu^*$ in the water area was approximately 0.07 [cm$^{-1}$] and the mean value of the reconstructed image including the air area was 0.026 [cm$^{-1}$]. Other two phantoms with the different radius also gave the same result (see Table S-1 in supplemental file).

It was found that the adaptive convergence parameter and the initial image setting
accelerated the convergence approximately three times faster than that with the constant convergence parameter starting with the ``zero" image.

\begin{figure*}[t]
\begin{center}
\includegraphics[width=10cm]{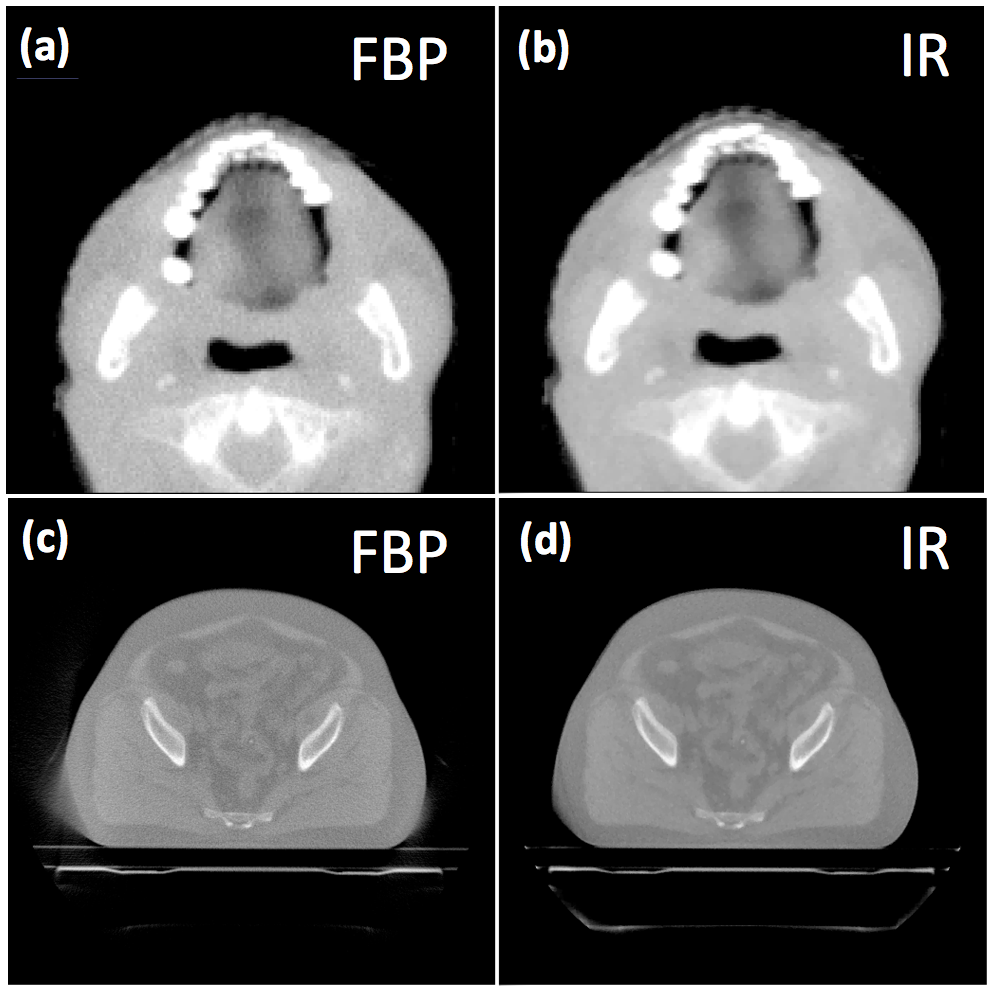}
\end{center}
\caption{Reconstructed images: (a) head region using filtered back projection, (b) head region using iterative reconstruction, (c) abdominal region using filtered back projection, and (d) abdominal region using iterative reconstruction.}
\label{fig5}
\end{figure*}

\subsection{CPU parallelization and OSEM method}

The convergence behavior with applying the OSEM method is also shown in Fig.~3, where the oscillation behavior is emphasized.
The convergence needs relatively larger iteration steps in the OSEM method, because the reconstructed image and its reprojection are obtained from thinned-out projection data. However, this apparent disadvantage is overcome by large reduction of the calculation time for the one iteration, thanks to the thinning. Eventually, the OSEM method highly accelerates the reconstruction speed.
We found that 500 iterations were an enough number of iteration in which the square of difference converges in the case of the thinning parameter more than $1/20$, whereas more iterations was required in  the thinning parameter of $1/40$.
Combining OSEM with CPU parallelization yielded the image reconstruction time
more than $40$ times faster than the original single CPU implementation, and more than $8$ times faster than the CPU parallelization without OSEM.
Consequently, combining the OSEM with CPU parallelization considerably accelerates the calculation.

\subsection{GPU computing}

In our reconstruction algorithm, the reprojection and the estimation processes occupy a large fraction of the calculation time.
Therefore, we coded these parts of the reconstruction using CUDA GPU processing.
Table I shows the average calculation time for reconstructing one image slice.
Although the results depend on the iteration number and OSEM parameter, the GPU code produces an image more than $10$ times faster than the parallelized CPU code without degrading image quality.
Using the GPU code, a high quality image can be reconstructed at a speed of $3.71$ seconds per slice.
This large improvement in iterative reconstruction speed on the GPU is achieved without any degradation in image quality compared to the CPU reconstructed images.

\begin{table}[htb]
\caption{Reconstruction speed [s] per one slice image with $512\times512$.}
\begin{tabular}{cccccccc}
\hline
\hline
Thinning parameter & 1 & 1/2 & 1/5 & 1/10 & 1/20 & 1/40\\
\hline
CPU (12 threads)   & 442.91 & 244.34 & 138.77 & 80.68 & 58.79 & 43.29 \\
CPU (20 threads)   & 325.40 & 191.16 & 94.40   & 51.21 & 35.99  & 24.59 \\
GPU (GTX1080Ti)  & 11.00   & 7.76 & 5.37 & 4.30 & 3.71 & 3.39 \\
\hline
\hline
\end{tabular}
\end{table}

\subsection{Reconstructed images}

Finally, the representative reconstructed images with $\alpha = 0.0002$ are shown in Fig.~4. For comparison, FBP reconstruction images are also shown there. In addition to the noise reduction seen in Figs.~4(a)-4(b), a relatively large reconstructed region (field of view) was achieved with the iterative reconstruction approach seen in Figs~4(c)-(d). 
Owing to the fan beam geometry, information of the projection data decrease at a fringe region of FOV.
Therefore, FBP fails to reconstruct the image at the edge of the body (Fig.~4-(c)).
On the other hand,
the statistical iterative reconstruction algorithm can reconstruct the image with a fewer projection data (the examination of the reconstruction using a half of projection data is given in supplemental file), and thus shows much better image quality at the edge (Fig.~4-(d).
From the viewpoint of the image quality, it can be said that there is a large benefit to use the iterative reconstruction algorithm clinically.

\section{Conclusion}

We investigated the clinical feasibility of a statistical iterative reconstruction method for MVCT on a HT system.
An image from the iterative reconstruction can be obtained within few seconds per image slice with a size of $512\times512$.
In addition, the image quality of our iterative method is much better than that of the FBP.
It can be concluded that the iterative reconstruction with GPU is fast enough for clinical use, and largely improves the MVCT images.
\\

\noindent
{\bf Conflict of interest}\\
Edward Chao and Calvin Maurer are officers of Accuray Inc.\\

\noindent
{\bf Funding}\\
Keiichi Nakagawa has a research grant from Accuray Inc.\\

\noindent
{\bf Ethical approval}\\
The study was ethically approved by the institutional review board at the University of Tokyo Hospital (reference number 3372).\\

\end{document}